# Review on the Role of Virtual Reality in Reducing Mental Health Diseases Specifically Stress, Anxiety, and Depression

Sadia Saeed[a], Khan Bahadar Khan[b], Muhammad Abul Hassan[c], Abdul Qayyum[d] and Saba Salahuddin[a]

[a]*Department of Electrical Engineering, Faculty of Engineering and Technology, The Islamia University, Bahawalpur, Pakistan*
[b]*Department of Information and Communication Engineering, Faculty of Engineering and Technology, The Islamia University, Bahawalpur, Pakistan*
[c]*Department of Biomedical Engineering, NED University of Engineering and Technology, Karachi, Pakistan*
[d]*National Heart and Lung Institute, Imperial College London, London, UK*

## ARTICLE INFO

*Keywords*:
Virtual Reality (VR)
Stress
Anxiety
Depression

## ABSTRACT

Objective: Virtual Reality (VR) is a technological interface that allows users to interact with a simulated environment. VR has been used extensively for mental health and clinical research. Mental health disorders are globally burdening health problems in the world. According to the Psychological Interventions Implementation Manual published by WHO on 6th March 2024, around one in eight people in the world lived with a mental disorder. This literature review is synthesized to find out the effects of VR therapy on stress, anxiety, and depression. Method: We used the Google Scholar database using the keywords VR, stress, anxiety, and depression. Publications from the last ten years (2014 to 2024) are considered. Only research papers and articles in the English language are included. All the research papers and articles with the keyword VR missing were rejected. Result: Google Scholar yielded 17,700 results from our keywords. Nine studies met our search criteria that are included in this review. Out of nine, five studies encountered mental stress and gave effective results in reducing it by VR therapy. The other four targeted mood disorders, social anxiety disorders, depression, loss of happiness, and sleep deprivation. They also showed immense potential in reducing mental illness while using VR. Conclusion: Findings are in favor of the effectiveness of VR in reducing stress, anxiety, and depression. Still, there is insufficient evidence to consider VR as a solely independent treatment over traditional medication. In the future, the limitations can be overcome by relying on VR and using it in hospitals as a reliable source of cure for mental illness.

## 1. Introduction

In diseases worldwide, mental disorders have significant burdens and separate contributions. Keiling et al.(1) wrote in their cross-sectional study about Worldwide Prevalence and Disability From Mental Disorders Across Childhood and Adolescence, among youth and children, there is a high chance that every one individual out of 10 (or approx. 293 million) with age between 5 to 24 years worldwide lives with a diagnosable mental disorder. Around a fifth part of all diseases linked with disabilities were credited to mental disorders in this population. The utmost neuropsychological conditions that come up with disability-adjusted life years (DALY), estimates the burden of disease globally launched by the World Bank, are mental health disorders (2). Due to the extreme pressure of work in today's technologically advancing era, an individual's mental and physical health is deteriorating which is another leading cause of high stress levels (3). A person can experience serious physical health issues like stomach aches, heart diseases, diabetes, and even cancer if mental disorders remain untreated for a long time (4) (5). Stress, Anxiety, and Depression are focused on mental health disorders in this review.

### 1.1. Prevalence of Mental Health

The comorbidity between these mental diseases among adults was studied in Ghana. A total data of 2456 adults were taken from the UHAS-Yonsei University Partnership Project. The total percentage of normal adults was found to be 48.2% whereas 51.8% lie in some mental health condition. The match between mental health issues is 25.2%, 53.3%,

sadiasaeedkhan6793@gmail.com (S. Saeed); kb.khattak@gmail.com (K.B. Khan); abulhasan@cloud.neduet.edu.pk (M.A. Hassan); a.qayyum@imperial.ac.uk (A. Qayyum); sabasalahuddin39@gmail.com (S. Salahuddin) and 9.7% for depression, anxiety, and stress respectively as shown in Figure 3. Hubert concluded in this study that there are elevated comorbidities of mental health issues among adults. Among numerous global interventions, to attain Sustainable Development Goals (SDGs), the country has to work on its Mental Health Act and should implement a Mental Health fund effectively to achieve SDGs by 2030 (6).

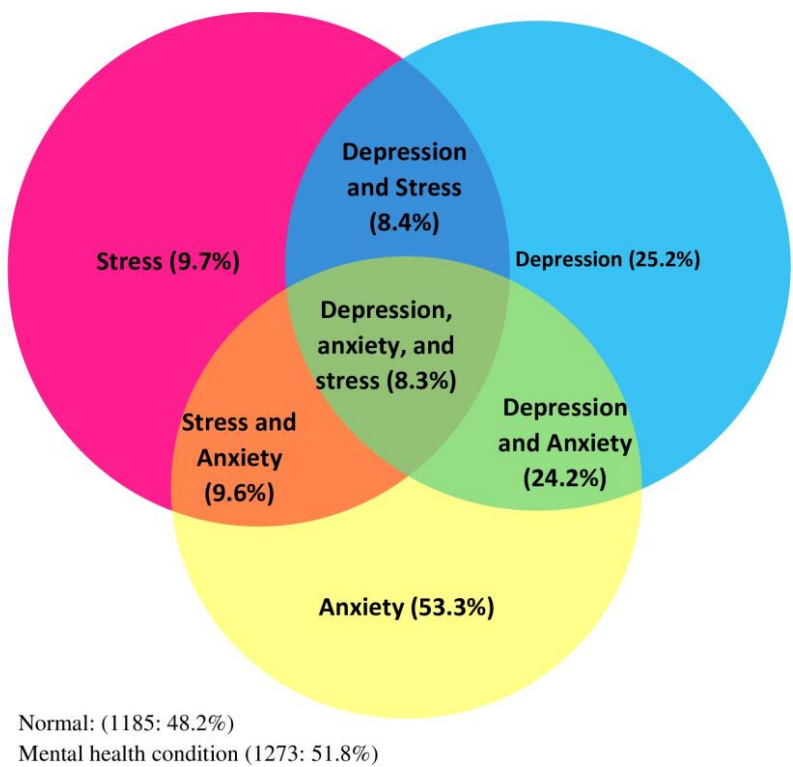

- Normal: (1185: 48.2%)
- Mental health condition (1273: 51.8%)

**Figure 1:** Prevalence of depression, anxiety, and stress. (6)

## 1.2. Stress

The initial idea of stress was discovered by Hans Ley in 1935. It was identified as a syndrome while experimenting on rats in a laboratory. However, this theory was rejected by physiologists until the 1970s. Later on, Seley's concept was universally recognized by physiologists in 1990 (7). She defined stress as a non-specific response to various requirements of the human body, plants, animals, and even bacteria. Seley professed that stress can't be separated from the daily life of human beings because it is related to such factors as emotional trauma, surgical trauma, blood loss, pain, fear, fatigue, workload, and frustrations that people are struggling with on a daily basis (8). Stress can cause other mental health problems such as anger, frustration, and depression which can lead to even suicidal thoughts (4) (9). In 1976, Seley differentiated stress into two categories- eustress and distress. Eustress is referred to as positive stress which results in growth in the working ability and capacity of an individual giving a positive response to the environment. It is considered as beneficial stress. On the other hand, if this stress is no longer motivating but its repetition becomes unable to handle then it will be called distress (7).

## 1.3. Anxiety

An individual can be in a mental state that evoke a feeling of threat in him which is referred as anxiety. It is very normal in human beings and a part of their daily life but can become an illness if overdue. As a result of this feeling, the psychological behavior of a person can be response to protect himself from danger. These psychological behaviors are part of the universe's mechanism that help organisms adapt to unfavorable environmental conditions. Anxiety can be classified broadly into two types' state and acute anxiety. State anxiety is a temporary response to the feeling of tension, like a person getting late from work can experience state anxiety while acute anxiety is a long-term condition in which a person may experience episodes of panic attacks (10). Diagnostic and Statistical Manual of the American

Psychiatric Association (11) classified anxiety disorders into six categories: simple phobia, social phobia (also called social anxiety disorder), generalized anxiety disorder, post-traumatic stress disorder (PTSD), panic disorder, and obsessive-compulsive disorder (OCD). Alone United States' 20 percent population is getting affected by these disorders costing $44 billion annually (12).

### 1.4. Depression

The Global Burden of Disease (GBD) study, initiated in 1992 in collaboration with the World Health Organization (WHO), identified major depressive disorder (MDD) as the fourth leading cause of disability and predicted that by 2020, it would rise to the second rank in terms of severity (13). It's a disorder likely be life-threatening at any age affecting more than millions of people from their childhood to old age. Recent surveys conducted on depression revealed that its occurrence in the lifespan of overall population ranges from 10% to 15%. Even if depressive disorders are treated and cured, they still burden a person throughout their lifetime because altogether symptoms rarely go away with the risk of reoccurrence (14). George S (15) classified depressive disorders in geriatric people into five: Major depressive disorder, minor depressive disorder, Dysthymic disorder, Bipolar disorder and Adjustment disorder with depressed mood.

### 1.5. Treatment of Stress, Anxiety, and Depression

Following are some treatments for stress, anxiety and depression extracted from the literature.

#### 1.5.1. Physical Activity

A study published in 2024, done on teachers, about using physical activity as a mediator in affecting levels of stress, anxiety, and depression. They divided the participants into two groups by engaging one group in physical activity for more than 150 minutes weekly and the other one was performing less than 150 minutes. They evaluate them based on some questionnaires including, the Psychological Well Being (PWB) Scale, Depression, Anxiety and Stress Scale (DASS- 21), and Socio-Demographic questionnaire. Results showed significantly lower levels of stress, anxiety, and depression in the participants who performed physical activity for more than 3 hours weekly (16). Studies show that an increase in physical activity helps to decrease the effects of negative thoughts and emotions. It elevates a better mood for over several hours (17). De Bruijn claims that when a person is physically active, his brain releases some neurotransmitters due to the activation of some brain areas. He suggested intense sports practice for 5 to 30 minutes daily in a consistent manner (18).

#### 1.5.2. Nutrition and Diet

Some researchers drew attention to the correlation between nutrition and mental health. It is an important factor that can prevent and help in treating mental disorders. Diet can play an efficacious role in decreasing depressive disorders and improving a healthy lifestyle. A healthy diet should contain essential micronutrients like zinc, magnesium, Vitamins B, C, E, and particularly n-3 fatty acids for reducing anxiety (19). Another study carried out a comparative study on university students of Ondokuz Mayıs University, Turkey. Students were put on a Mediterranean diet for one month. It's a plant-based diet that recommends the consumption of vegetables, fruits, cereals, and lentils daily, and white meat weekly. The use of olive oil as a lipid source is also recommended. Consumption of sweets and chocolates is discouraged. After the statistical analysis based on the Mediterranean Diet (MD), quality index for children and adolescents (KIDMED), Pittsburgh sleep quality index (PSQI), and Depression anxiety stress scale (DASS-42), they concluded that MD proved to be significant in decreasing depression scales (20).

#### 1.5.3. Vitamin D

Karisa Renteria wrote about the association of Vitamin D (VD) with anxiety and depressive disorders. Ultraviolet from sun exposure is the source of VD synthesis primarily in the human body. In Northwest China, patients suffering from rheumatoid arthritis faced depression and anxiety according to the Hamilton Depression and Anxiety statistical scales (21). Lab experimentation on rodents proved that VD repeated supplementation decreases depression and has an effect like anti-depressants (22).

#### 1.5.4. Music therapy

Research indicates that music enhances symptom control, uplifts mood, alleviates anxiety, and enhances overall quality of life. A study published recently in 2024, in which 30 women who were patients undergoing brachytherapy were exposed to music intervention. Results showed a significant decrease in depression (23). Rossetti et al. (24), in their research on 78 cancer patients significantly lowered stress levels with music therapy showing its promising role in reducing stressors. Chen et al. (25) examined the impact of listening to a brief 15-minute selection of music before radiation therapy (RT), where patients were able to choose music tracks from various genres (pop, traditional, classical). The study found that this musical intervention led to reduced levels of both state-trait anxiety and systolic blood pressure.

### 1.5.5. Electrical stimulation

Brain stimulation through non-invasive techniques allows distinct opportunities in both healthy individuals and those with clinical conditions to not just investigate brain functions but also to alter fundamental physiological aspects of human behavior and cognition. Alizadehgoradel et al. (26) experimented on Obsessive Compulsive Disorder (OCD) patients with electrical stimulation. The intensified stimulation of 2-mA proved the therapeutic importance of electrical stimulation.

### 1.5.6. Virtual Reality

Virtual Reality is an immersive digital technology that individuals find exciting by experiencing artificially created environments. During the past decade, VR has being used for rehabilitation therapy purposes as behavioral medication to treat many mental illnesses including Post-traumatic stress disorder (PTSD), phobias, and autism (27). The use of virtual reality (VR) treatments for various mental health issues has reported positive results. Recently VR exposure therapy has become popular in treating anxiety and depression, with increasing research in this area proving the efficacy of VR treatment. It is being recognized as a successful method for addressing symptoms associated with anxiety and depression (28).

VR refers to a 2D or 3D environment generated by a computer. The simulation consists of relaxing images like a forest and soothing instrumental sounds in the background that depict real situations. A person using the equipment feels the whole scenario as real or in a physical way (29).

Generally, the basic components of the VR system are shown in Figure 1. The figure is based on the keyword VR attached to the VR system which is comprised of hardware plus software including the physical environment and the user altogether integrated into a computer-generated VR environment. LaValle (2019) highlighted the essential role of the user in a VR system as the sensors integrated into the system perceive response from human vision and motions in real time that is why user immersion is needed. Including the real physical world in a virtual reality environment, perception is also necessary as there can be certain stimuli like auditory, tactile, or momentary observed by the user. When all these required components are unified, the user experiences the Virtual Environment (VE) and starts accepting a fully computer-generated environment (30). This feeling can be different from the real world due to changes in space, scales, position, and perception (31). These described components are publicly available in the market and are expected to be extensively in demand due to a reduction in cost and rapid growth in performance (32).

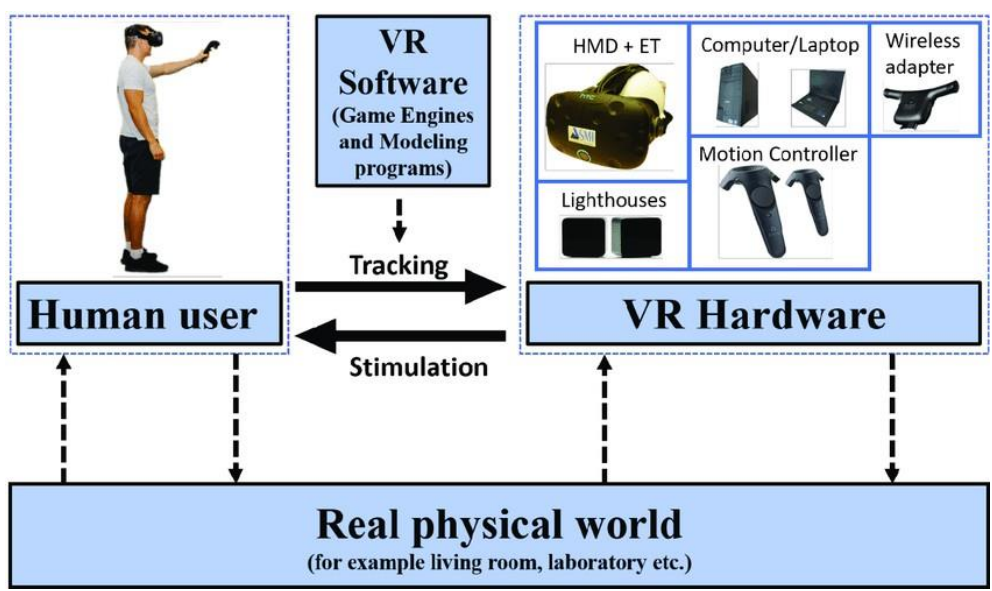

**Figure 2:** General components of VR (30)

## 1.6. VR and Mental Health

VR systems are being applied in the medical field including many disciplines mainly as psychotherapy for relieving mental stress, social anxiety disorders, mood disorders, sleep deprivation Disorders, etc. and it is proven to be restorative when combined with the natural environment (33) In 1990's, research on college students with acrophobia was conducted by Rothbaum et al. (34) and the authors concluded the potency and effectiveness of VR in reducing fear of heights. A study done in 2010 revealed it's an effective therapy in reducing symptoms of post-traumatic stress disorder (PTSD) in military personnel who have a history of combat-related traumas (35).

## 2. Methodology

### 2.1. Search Strategy

We focused on strategy explicitly fit to search mostly relevant papers to VR. We concentrated on the outcomes of VR usage to find its role in support of mental health issues.

### 2.2. Databases Searched

Google Scholar is one of the easiest and quickest ways to access a vast number of research papers, journals, and databases that we used as the main source in this research. We defined specific eligibility criteria to limit our research landscape.

### 2.3. Search Terms

Our keywords were Virtual Reality (VR), Stress, Anxiety and Depression. The acronym VR is used sometimes to represent the term Virtual Reality but we used the full term to avoid any clashes with other acronyms of VR (e.g. Voice recognition).

### 2.4. Eligibility Criteria

As technology is rapidly evolving progressively in the field of VR, we decided to restrain our publication year from 2014 to 2024. Google Scholar facilitates different custom ranges to search articles. With our keywords, Google Scholar gave 17,500 results. After screening and analysis, 9 papers were selected and studied. Exclusion criteria were non-English language papers, no use of VR headset equipment and not targeting stress, anxiety, and depression particularly. The flowchart in Figure 3 shows hierarchy of selection criteria.

## 3. Results

### 3.1. Subjects sorting

The first step is the gathering and sorting of subjects according to the goal of the investigation. A minimum number of subjects selected is n=15 (36) and the maximum is n=48 (37). Out of nine, six researchers chose healthy participants, one took elderly participants from nursing homes, and two selected patients with major Social Anxiety Disorder (SAD) (38), major depressive disorder (MDD), and bipolar disorder (BD) (39).

### 3.2. Key Findings

After review of selected papers, we put together descriptive characteristics and mentioned important aspects for easy comparison in Table 1.

1. Kamaniska (40) did two researches in 2020 and 2021. In 2020, he did six sessions of VR therapy on 23 healthy office workers. Each session consists of 63 seconds. With the combination of EMDR, a psychotherapy treatment proved to be a successful reducer of stress and anxiety. The VR scenario he used is a 2D forest environment with an instrumental background (40). The first scene was to calibrate the scene with the alignment of participants horizontally and vertically. Sometimes the user gets dizzy through VR sessions if have cyber-sickness which can be called VR sickness (45). So to get familiar, the second session was designed for an adaptation to VR in which a person can walk in a forest environment with the background sound of chirping birds. The task was to follow a golden ball that was moving horizontally in the environment. Pre and post-questionnaire was connected electronically with the VR system to avoid any inconsistency. The psychological measurements consist of questionnaires GHQ-12 (12-Item General Health Questionnaire), GMS (General Mood Scale), ASL (Actual Stress Level) and VRSQ Virtual Reality Sickness Questionnaire). Data observed from the assessment showed the significance of VR therapy in reducing measured stress. Elevating moods of participants were detected after relaxation sessions. The use of EMDR with VR therapy turned out to be excellent in reducing mental stress (40).
2. Kamaniska in 2021, did VR therapy on 28 healthy subjects but with three sessions for 240 seconds. The scenario was the same forest environment. In this work, he did a behavioral and neurological assessment. In behavioral assessment, he did three Stroop tests in between VR sessions (41). Stroop test consists of some tasks and is used as a stressor in this research. In the Stroop test person has to identify the ink color of the printed word rather than read the text (E.g. the word "blue" is printed with the color "orange"). This confuses a person and puts stress on the brain making him answer slowly (46). In neurological assessment, electroencephalography (EEG) was taken all the time during the Stroop test and VR therapy. The EEG system used was 32 channels with 16 electrodes which recorded brain wave activity. During the Stroop Test in the starting minutes, changes in the alpha wave were observed in 18 participants that showed an increase in stress level induced purposely. While giving VR with EMDR, the gradual damping of beta waves could be seen in 1/3 of the participants. It shows the relaxation stage (41).

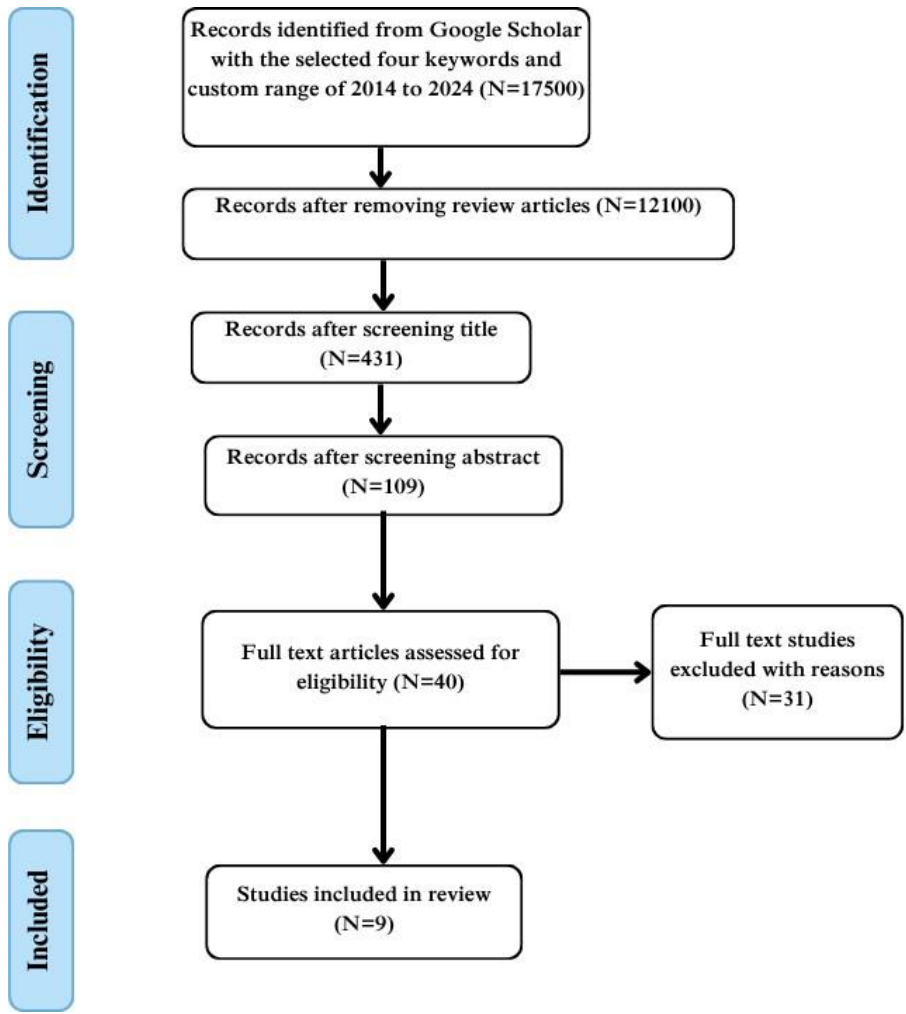

**Figure 3:** Flow diagram of studies selection

**Table 1**
Characteristic illustration of Included Studies

| S.no | Ref. | Year | Subjects | Behavioral Assessment | Psychometric Assessment | Neurological Assessment | VR Therapy Type | Duration |
|---|---|---|---|---|---|---|---|---|
| 1. | (40) | 2020 | N=23 Healthy Office workers m=10, f=13 | - | i. GHQ-12 ii. GMS iii. ASL iv. VRSQ v. TS | - | Forest scenery 2D, EMDR, Instrumental background | 63s x 6 |
| 2. | (41) | 2021 | N=28 Healthy office workers m=19, f=9 | Stroop | - | 32 channel EEG 16 electrodes | Forest scenery 2D, EMDR, Instrumental background | 240s x 3 |
| 3. | (42) | 2021 | N=23 Healthy volunteers | MIST | - | EEG system four electrodes Fp1, Fp2, F5, F6 | 3D scenarios aurora borealis a solitary beach a space trip | 5m x 1 |
| 4. | (43) | 2018 | Initially, N=65 healthy students selected N=20 | Stroop | DASS-21 | EEG with RMS SuperSpec 10-20 electrode | 2D forest stroll facility natural sounds | 10m x 1 |
| 5. | (44) | 2017 | N=20 healthy participants m=14, f=6 age=18 to 21yr | i. Stroop ii. GoNogo Visual (GNG) task x 2 | i. DASS-21 ii. PANAS | - | 2D relaxing island landscape | 10m x 1 |
| 6. | (37) | 2020 | N=60 elderly participants nursing homes | | i. OHQ ii. PSS iii. PSQI iv. EOM-DM v. life satisfaction | - | 3D VR Hands-On Aromatherapy | 2Hr x 9 |
| 7. | (36) | 2021 | N=15 University students | - | i. BAI ii. BDI-II iii. PSS | - | iVR exergame intervention | 30m x 12 |
| 8. | (38) | 2021 | N=28 SAD Patients m=3,f=19 N=27 Healthy Controls (HCs) m=14 , f=13 | - | i. BAI ii. STAI-X iii. ISS iv. PERS v. K-SPS vi. K-SIAS vii. BFNE viii. K-SAD ix. LSAS | fNIRS system 48 channels | 3D scenario social situations with audience | 30s x 6 |
| 9. | (39) | 2015 | N=22 Patients with MDD and BD M=6, f=16 age=21 to 65yr | - | i. DASS-21 ii. PRS iii. KSSMQ | - | Visual presentations relaxing music | 1hr x 3 |

3. In a study done by Perez-Valero in 2021, 23 subjects were taken who volunteered with a selection criteria of no physical or mental health issues. He did VR therapy with behavioral and neurological assessment. The experiment consists of a Montreal imaging stress task (MIST) as a stressor followed by 3D VR. Participants were also equipped with an EEG system during the entire session for recording EEG activity. A versatile EEG system was used with 4 electrodes Fp1, Fp2, F5, and F6 (42). MIST is an array of computerized mathematical tasks like a combo of addition, multiplication, subtraction, and division (47). In this research, the goal was to predict stress levels through EEG during VR while VR was used as a relaxer. The study successfully identifies the stress levels but this approach is feasible only to detect two to three classes' i.e. low, medium, and high stress.

4. In the fourth study, initially, 65 students of the biomedical department were taken as subjects. Later on, they filtered them with DASS-21 test scores and ended up with 20 subjects divided into two equal groups, stressed and mildly stressed (43). Depression, Anxiety, and Stress Scales (DASS-21) is a psychometric assessment test consists a 21-item questionnaire. It can assess distress, lacking interest, state of anhedonia, lack of happiness, and depression (48). In behavioral assessment, he used the Stroop test and the Go Nogo task (mathematical task) as stressors. In neurological assessment, EEG was done during VRT for ten minutes as a relaxation therapy just similar to in the study (41). Studies showed improvement in the performance of tasks after VR therapy and EEG obtained depicted increasing calmness after therapy (43).

5. In 2007, Taneja did a study of 20 participants divided into two groups, control and stressed, based on DASS-21 score. He used the Stroop test and Go-Nogo (GNG) task as stressors and VR therapy for ten minutes as a relaxer. After the session, participants were directed to fill Positive Affect Negative Affect Scale (PANAS) questionnaire as a psychometric assessment [48]. PANAS is 20-item questionnaire that indicates the person's two mood scales positive and negative (49). PANAS results indicate the effectiveness of VR therapy to reduce mental stress (44).

6. From a nursing home, 60 elderly people were selected as participants over the age of 65 years. They were divided into two groups, the experimental and the control group. The experimental group received 3D VR therapy with the combination of aromatherapy for 2 hours weekly till 9 weeks by certified Aromatherapists from the US and UK (37). The use of essential oils extracted from aromatic plants and herbs for healing purposes is called Aromatherapy. In a review, six studies proved a significant reduction in anxiety by the use of aromatherapy (50). Psychometric assessment tests included the Oxford Happiness Inventory (OHQ), Perceived Stress Scale (PSS), Pittsburgh Sleep Quality Index (PSQI), Experiences During Meditation (EOM-DM), and life satisfaction scale. OHQ is a 29-item multiple-choice questionnaire to assess the happiness of a person on a Likert scale of six points. The higher the score higher the level of happiness (51). PSS is a 14-item questionnaire, 7 positive and 7 negative, that measures stress levels on a scale of 0 to 4 (52). PSQI is an effective measure of sleep quality in aged people. It measures seven domains from poor to good sleep (53). EOM-DM has originally five subscales: cognitive effects, emotional effects, mystical experiences, relaxation, and physical discomfort. Only initial two scales are used in this study to measure the experience of meditation on a Likert scale of 1 to 5 with good experience on the higher number side (54). Health Promotion Administration of the Ministry of Health and Welfare developed a life satisfaction scale for elderly people. It's a 20-item questionnaire but in this study, a short version consisting of only 10 items is used. Higher scores indicate higher life satisfaction (55). He concluded from these five variables that the combination of VR and Aromatherapy yielded an increase in happiness, a decrease in perceived stress, improved sleep quality, a good meditation experience and increased life satisfaction (37).

7. On fifteen university students, an experiment was done to assess the ability of iVR (non-immersive Virtual Reality) exergames to reduce anxiety, stress, and depression. Students needed to play FitXR (FitXR Limited) game twice per week for 6 weeks. Game play was of 30 minutes per session. Three psychometric tests were done before and after VR sessions, Beck Anxiety Inventory (BAI), Beck Depression Inventory-II (BDI-II), and Perceived Stress Scale (PSS) (36). BAI and BDI-II both are 21-item questionnaires that measure anxiety and depression levels (56). The author concluded the results that iVR games have the potential and good usability in reducing depression with no significant difference in anxiety and stress (36).

8. In Korea, a study on 28 Social Anxiety Disorder (SAD) patients was done to find out the effectiveness of VR on SAD patients using functional near-infrared spectroscopy (fNIRS). 6 sessions of 30 seconds were provided to patients. VR scenario was comprised of social situations and audience. Patients had to introduce themselves in front of them. To evaluate, psychometric and neurological assessment took place (38). To assess social anxiety, the Korean version of the Beck Anxiety Inventory (BAI) (57) and State-Trait Anxiety Inventory-X (STAI-X) (58), Internalized Shame Scale (ISS) (59), the Post-Event Rumination Scale (PERS) (60), the Korean version of the Social Phobia Scale (K-SPS) (61), Korean version of the Social Interaction Anxiety Scale (K-SIAS) (61), Brief- Fear of Negative Evaluation (BFNE) scale (62), the Korean Social Avoidance and Distress (K-SAD) scale (63) and the Liebowitz Social Anxiety Scale (LSAS) (64). For neurological assessment, fNIRs of the brain's pre-frontal cortex region with 48 channels were observed. fNIRs of the brain is a spectroscopy technique to find out the functional activity of brain tissue by measuring concentration changes in oxygenated hemoglobin and deoxygenated hemoglobin. It's a safe, user-friendly, and portable non-invasive method as compared to other brain imaging techniques like functional MRI (fMRI) (65). Results of SAD patients and Healthy controls were compared. VR therapy was found to be effective in reducing symptoms in SAD patients (38).
9. In Singapore, a study on people with mood disorders was done to investigate the effects of VR-based programs in reducing their symptoms. On 22 patients diagnosed with Major Depressive Disorder (MDD) and Bipolar Disorder (BD), VR therapy was done for three days for 1 hour per session. VR programs consist of relaxing videos and soothing background music with auditory instructions in a soft voice. Pre and post-psychometric assessment tests, DASS-21, PRS, and KSSMQ, were observed to compare the effectiveness of VR. Depression, anxiety, and stress scale (DASS-21) questionnaire was evaluated on 4 point scale (0 to 3). Participants rated the Perceived Relaxation Scale (PRS) on a ten-centimeter scale from zero (very tense) to ten (very relaxed). Knowledge on Stress and Stress Management Questionnaire (KSSMQ) was a 15-item questionnaire designed by researchers to estimate the knowledge of patients on scores 0 to 15. Higher scores indicated higher knowledge. All three variables measured and patients interviewed depicted positive effects of VR in reducing mood disorder symptoms (39).

## 4. Conclusion

Table 2 illustrates the effects of VR therapy in each research included in the review along with the mental disorder encountered.

**Figure 4:** Concluded Effects

| S.no. | Ref. | Mental Stress | Anxiety | Depression | Happiness | Sleep Quality | Mood Relaxation | Social Anxiety Disorder |
|---|---|---|---|---|---|---|---|---|
| 1. | (40) | ↓ | ↓ | - | - | - | ↑ | - |
| 2. | (41) | ↓ | - | - | - | - | ↑ | - |
| 3. | (42) | ↓ | - | - | - | - | ↑ | - |
| 4. | (43) | ↓ | - | - | - | - | ↑ | - |
| 5. | (44) | ↓ | - | - | - | - | ↑ | - |
| 6. | (37) | ↓ | - | - | ↑ | ↑ | ↑ | - |
| 7. | (36) | / | / | ↓ | - | - | - | - |
| 8. | (38) | - | - | - | - | - | - | ↓ |
| 9. | (39) | ↓ | ↓ | ↓ | - | - | ↑ | - |

The principle focus of this research was to review concisely accessible literature regarding the effectiveness of VR therapy in improving mental health and reduction of mental disorders mainly focusing on stress, anxiety, and depression. Nine studies were finalized and included in this review study. Key findings disclosed immense improvement. VR therapy proved to be a great source of mood relaxation in seven studies. Any unfavorable effects of VR therapy were not found. Comprehensively, this review including nine studies presents an initial verification of VR therapy that it has the potential to reduce stress, anxiety, and depression.

## 5. Limitations

While interpreting the results of this review, some limitations must be considered. To obtain accurate results, we used the Google Scholar database for this study, which may have excluded other databases containing relevant articles. The study was limited to papers published in the last ten years. We focused only on three keywords—stress, anxiety, and depression—to narrow the search results. Other mental health issues related to these, such as phobias and PTSD, should also be considered when evaluating the effects of VR therapy. Additionally, only articles published in English were included. Due to the large number of search results, some relevant publications may have been overlooked.